\documentclass[11pt]{article}
\usepackage{fullpage}

\usepackage{amsmath}
\usepackage{graphicx}
\usepackage{amsfonts}
\usepackage{amssymb}

\newcommand{\one}{\leavevmode\hbox{\small1\normalsize\kern-.33em1}}
\def\tr{\mbox{tr}}

\def\be{\begin{equation}}
\def\ee{\end{equation}}
\def\ben{\begin{eqnarray}}
\def\een{\end{eqnarray}}
\def\eea{\end{array}}
\def\bea{\begin{array}}

\newcommand{\bei}{\begin{itemize}}
\newcommand{\eei}{\end{itemize}}

\def\A{{\cal A}}
\def\B{{\cal B}}
\def\E{{\cal E}}

\begin{document}

\title{Secure device-independent quantum key distribution\\ with causally independent measurement devices}
\author{Llu\'{\i}s Masanes$^1$, Stefano Pironio$^2$, Antonio Ac\'in$^{1,3}$\\[0.5em]
{\it\small $^1$ICFO--Institut de Ci\`encies Fot\`oniques, E--08860 Castelldefels, Barcelona, Spain}\\
{\it\small $^2$Laboratoire d'Infomation Quantique, Universit\'e Libre de Bruxelles, 1050 Bruxelles, Belgium}\\
{\it\small $^3$ICREA--Instituci\'o Catalana de Recerca i Estudis Avan\c{c}ats, E--08010 Barcelona, Spain}}
\date{}

\maketitle

\begin{abstract}
Device-independent quantum key distribution aims to provide key distribution schemes whose security is based on the laws of quantum physics but which does not require any assumptions about the internal working of the quantum devices used in the protocol.
This strong form of security, unattainable with standard schemes, is possible only when using correlations that violate a Bell inequality.
We provide a general security proof valid for a large class of device-independent quantum key
distribution protocols in a model in which the raw key elements are generated by causally independent measurement processes. The validity of this independence condition may be justifiable in a variety of implementations and is necessarily satisfied in a physical realization where the raw key is generated by $N$ separate pairs of devices.
Our work shows that device-independent quantum key distribution is possible with key rates comparable to those of standard schemes.
\end{abstract}

\section{Introduction}
A central problem in cryptography is the distribution among distant users of secret keys that can be used, e.g., for the  secure encryption of messages. This task is impossible in classical cryptography unless assumptions are made on the computational power of the eavesdropper. Quantum key distribution (QKD), on the other hand, offers security against adversaries with unbounded computing power \cite{BB84}.

The ultimate level of security provided by QKD was made possible thanks to a change of paradigm. While in classical cryptography security relies on the hardness of certain mathematical problems, in QKD it relies on the fundamental laws of quantum physics. A side-effect of this change of paradigm, however, is that whereas the security of classical cryptography is based on the mathematical properties of the key itself --- how the key was actually generated in practice being, in principle, irrelevant to the security of the scheme --- in QKD, the security crucially depends on the physical properties of the key generation process, e.g., on the fact that the key was produced by measuring the polarization of a single-photon along well defined directions. But then, how can one asses the level of security provided by a real-life implementation of QKD, which will inevitably differs in inconspicuous ways from the idealized, theoretical description~\cite{scarani}? Errors in the encoding 
 of the signals of Alice~\cite{lo}, for instance, or features of the detectors not taken into account in the theoretical analysis~\cite{makarov} can be exploited to break the security of real-life QKD schemes.

Device-independent QKD (DIQKD)~\cite{aci07} aims at closing the gap between theoretical analyses and practical realizations of QKD by designing protocols whose security does not require a detailed characterization of the devices used to generate the secret key (such as, e.g., the dimension of the Hilbert space of the quantum signals or the type of measurements performed on them)~\cite{aci07,Mayers,AGM,pironio09}. This stronger form of cryptography is possible if it is based on the observation of a Bell inequality violation, which guarantees that the data produced by the quantum devices possess some amount of secrecy, independently of how exactly these data were generated~\cite{Ekert,BHK}. In some sense, DIQKD combines the advantages of classical and quantum cryptography: security against unbounded adversaries based on the law of quantum physics but which does not rely on the physical details of the generation process. A fully device-independent demonstration of QKD, however, 
 still represents at present an experimental challenge~\cite{ampli}.

In this work, we provide a general formalism for proving the security of DIQKD protocols. The key element in our analysis is a bound on the min-entropy of the raw key as a function of the observed Bell inequality violation.
Compared to the security proof given in ~\cite{aci07,pironio09,mckague}, which is restricted to protocols based on the Clauser-Horne-Shimony-Holt (CHSH) inequality~\cite{chsh}, our approach is completely general and can be applied to protocols based on arbitrary Bell inequalities. Furthermore, it is not limited to ``collective attacks'', but is valid against the most general attacks available to an eavesdropper.

The DIQKD model that we consider, however, is partly restricted as it supposes that the measurement processes generating the different bits of the raw key are causally independent of each other (though they could be arbitrarily correlated). This independence condition  is necessarily satisfied in a physical realization where the $N$ bits of the raw key are generated by $N$ separate pairs of devices used in parallel. Our analysis therefore shows that secure fully device-independent QKD is in principle possible. In a more practical realization in which a single pair of devices is used sequentially to generate the raw key, our measurement independence condition is satisfied if the devices have no internal memory, an assumption that may be justifiable in a variety of implementations. Note that our measurement independence condition and the level of security provided here is equivalent to the one considered in \cite{masanes,nonfinished,hrw}. The difference with respect to \cite{masanes,nonfinished,hrw} is that our proof does not rely only on the no-signalling principle but also on the validity of the quantum formalism. This results in much better key rates, comparable to those of standard QKD.

\section{General structure of a DIQKD protocol}
Let us start by presenting the class of protocols that we consider here, which are variations of Ekert's QKD protocol~\cite{Ekert,AMP}. Alice and Bob share a quantum channel that distributes entangled states and they both have a quantum apparatus to measure their incoming particles. These apparatuses  take an input (the measurement setting) and produce an output (the measurement outcome). We label the inputs and outputs $x$ and $a$ for Alice, and $y$ and $b$ for Bob, and assume that they take a finite set of possible values.

The first step of the protocol consists in measuring the pairs of quantum systems distributed to Alice and Bob. In most of the cases (say $N$), the inputs are set to fixed values $x_i =x_\mathrm{raw}$ and $y_i =y_\mathrm{raw}$ and the corresponding outputs  ${\bf a}= (a_1, \ldots a_N)$ and ${\bf b}= (b_1, \ldots b_N)$ constitute the two versions of the raw key. In the remaining systems, which  represent a small random subset of all measured pairs (of size  say $N_\mathrm{est}\approx\sqrt{N}$), the inputs $x,y$ are chosen uniformly at random. From these $N_\mathrm{est}$ pairs, Alice and Bob determine the relative frequencies $q(ab|xy)$ with which the outputs $a$ and $b$ are obtained when using inputs $x$ and $y$.  These relative frequencies quantify the degree  of non-local correlations between Alice and Bob's system through the violation of the Bell inequality associated to the DIQKD protocol. This Bell inequality is defined by a linear function $g$ of the input-output correlations $q(ab|xy)$:
\begin{equation}\label{g}
    g = \sum_{a,b,x,y} g_{abxy} q(ab|xy) \leq g_\mathrm{loc}\,,
\end{equation}
where $g_{abxy}$ are the coefficients defining the Bell inequality and $g_{\rm loc}$ is its local bound. A particular example of a Bell inequality is the CHSH inequality~\cite{chsh}
\begin{equation}\label{gchsh}
    g_\mathrm{chsh} = \sum_{a,b,x,y} (-1)^{a+b+xy} q(ab|xy) \leq 2\,,
\end{equation}
where $a,b,x,y\in\{0,1\}$.

After this initial ``measure and estimate'' phase, the rest of the protocol is similar to any other QKD protocol. Alice publishes an $N_\mathrm{pub}$-bit message about ${\bf a}$, which is used by Bob to correct his errors ${\bf b \rightarrow b'}$, such that ${\bf b' =a}$ with arbitrarily high probability. Alice and Bob then generate their final secret key ${\bf k}$ by applying a 2-universal random function to $\mathbf{a}$ and $\mathbf{b'}$, respectively \cite{2-universal}.

\section{The DIQKD model}
In the DIQKD approach, we do not assume that the devices behave according to predetermined specifications.
For instance, the state emitted by the source of particles may be modified by the eavesdropper, or
the implementation of the measuring devices may be imperfect. To analyze the security of a DIQKD protocol, we must therefore first specify how we model the $N$ pairs of systems used to generate the raw key.

These $N$ pairs of systems are eventually all measured using the
inputs $x=x_\mathrm{raw}$ and $y=y_\mathrm{raw}$, but since they
where initially selected at random and each of them could have
been part of the $N_\mathrm{est}$ pairs used to estimate the Bell
violation, we must also consider what would have happened for any
other inputs $x$ and $y$. Let therefore  $P(\bf{ab}|
\bf{xy})$ denote the prior probability to obtain outcomes ${\bf
a}$ and ${\bf b}$ if measurements ${\bf x}=(x_1, \ldots x_N)$ and
${\bf y}=(y_1, \ldots y_N)$ are made on these $N$ pairs. This
unknown probability distribution characterizes the initial system
at the beginning of the protocol.

In the theoretical model that we consider here, we view the $N$ bits of the raw key as arising from $N$ \textit{commuting} measurements on a joint quantum system $\rho_{\A\B}$.  That is, we suppose that the probabilities $P(\bf{ab}|\bf{xy})$ can be written as
\begin{equation}\label{multiP}
    P({\bf ab|xy}) = \tr[\rho_{{\A\B}}\, \prod_{i=1}^N A_i(a_i|x_i) B_i(b_i|y_i)]\,,
\end{equation}
where $A_i(a_i|x_i)$ are operators describing the measurements made by Alice on her $i^\text{th}$ system if she select input $x_i$ (they thus satisfy $A_i(a_i|x_i)\geq 0$ and $\sum_{a_i} A_i(a_i|x_i)=\one$), where, similarly, $B_i(b_i|y_i)$ are operators describing the measurements made by Bob, and where these measurement operators satisfy the commutation relations
\begin{equation}\label{commutator2}
[A_i(a|x),B_j(b|y)]=0
\end{equation}
and
\begin{equation}\label{commutator}
[A_i(a|x),A_j(a'|x')]=[B_i(b|y),B_j(b'|y')]=0
\end{equation}
for all $i,j$ and $a,a',b,b',x,x'$.
Apart from the conditions \eqref{commutator2} and \eqref{commutator}, the state $\rho_{AB}$ and the operators $A_i(a_i|x_i)$ and $B_i(b_i|y_i)$ are arbitrary and unspecified. The only constraint on them is that they should return measurement probabilities \eqref{multiP} compatible with the statistics of the $N_\mathrm{est}$ randomly selected pairs, characterized by the observed Bell-inequality violation $g$.

In quantum theory,
measurement operators that commute represent compatible measurements that do not influence each other and which can be performed independently of each other. The commutation relations \eqref{commutator2} between the operators $A_i(a_i|x_i)$ describing Alice's measurement devices and the operators $B_i(b_i|y_i)$ describing Bob's measurement devices are thus a necessary part of any DIQKD model; security cannot be guaranteed without them.

The commutation relations \eqref{commutator} between the operators $A_i(a_i|x_i)$ \textit{within} Alice's location, and the commutation relations between the operators $B_i(b_i|y_i)$ \textit{within} Bob's location, represent, on the other hand, additional constraints specific to the DIQKD model considered here. These commutation relations are satisfied in an implementation in which the $N$ bits of the raw key are generated by $N$ separate and non-interacting pairs of devices used in parallel.

In the extreme adversarial scenario where the provider of the devices is not trusted (e.g., if the provider is the eavesdropper itself), this independence condition can be guaranteed by shielding the $N$ devices in such a way that no communication between them occurs during the measurement process. One could also consider a setup where the measurements performed by the $N$ devices define space-like separated events. However, even in a space-like separated configuration, the ability to shield the devices is required if the provider of the devices is untrusted, as we cannot guarantee through other means that the devices do not send directly unwanted information to the adversary.  But, then, the ability to shield the devices is already sufficient by itself to guarantee \eqref{commutator}.

In a more practical implementation where the raw key is generated by repeatedly performing measurements in sequence on a \emph{single} pair of devices, the commutation relation \eqref{commutator} expresses the condition that the functioning of the devices should not depend on any internal memory storing the quantum states and measurement results obtained in previous rounds.
In the most general DIQKD model, the quantum devices could possess a quantum memory such that the state of the system after the $i^\text{th}$ measurement is passed to the successive round $i+1$ (this state could also contain classical information about the measurement inputs and outputs of step $i$). If $\rho^i_{\A \B}$ denotes the state of the system before measurement $i$, the unormalised state passed to round $i+1$ in the event that Alice and Bob use inputs $x_i$ and $y_i$ and obtain outputs $a_i$ and $b_i$ would then be $\tilde A^\dagger_i(a_i|x_i)\tilde B^\dagger_i(b_i|y_i)\rho_{\A \B}^i\tilde A_i(a_i|x_i)\tilde B_i(b_i|y_i)$ where $\tilde A_i(a|x)$ and $\tilde B_i(b_i|x_i)$ are generalized measurement operators describing Alice's and Bob's measurements and satisfying $\sum_{a} \tilde A_i(a|x)\tilde A_i^\dagger(a|x)=\sum_{b} \tilde B_i(b|y)\tilde B_i^\dagger(b|y)=I$. In such a model, the probabilities $P({\bf ab|xy})$ are then given by

\begin{equation}\label{gmod}
P({\bf ab|xy}) = \tr[\prod_{i=N}^{1} \tilde A^\dagger_{i}(a_i|x_i) \tilde B^\dagger_{i}(b_i|y_i)
   \, \rho_{{\A\B}}\, \prod_{i=1}^N \tilde A_i(a_i|x_i) \tilde B_i(b_i|y_i)]\,,
\end{equation}
where $\rho_{\A\B}$ denotes the initial state at the beginning of the protocol, and the order in the products is relevant. Imposing commutation relations between all operators pertaining to different rounds corresponds to neglect the causal order in \eqref{gmod} due to memory effects. We then recover a model of the form \eqref{multiP} by defining $A_i(a|x)=\tilde A_i(a|x)\tilde A_i^\dagger(a|x)$ and $B_i(b|y)=\tilde B_i(b|y)\tilde B_i^\dagger(b|y)$.

\section{Security Proof}
We now establish a bound on the secret key rate that can be achieved against an unrestricted eavesdropper Eve for a QKD protocol satisfying the description \eqref{multiP}, \eqref{commutator2}, \eqref{commutator}. The information available to Eve can be represented by a quantum system that is correlated with the systems of Alice and Bob. We denote by $\rho_{\A\B\E}$ the corresponding $(2N+1)$-partite state, with $\mathrm{tr}_\E\,\rho_{\A\B\E}=\rho_{\A\B}$.
This state describes the $2N+1$ systems at the beginning of the protocol. After the $N$ systems of Alice have been measured, the joint state of Alice and Eve is described by the classical-quantum state
\be \label{ccq}
\rho_{{\A}{\E}}=\sum_{\mathbf{a}} P(\mathbf{a}|\mathbf{x}_\mathrm{raw}) |\mathbf{a}\rangle\!\langle\mathbf{a}|\otimes  \rho_{\E|\mathbf{a}}\,,
\ee
where $\rho_{\E|\bf{a}}$ is the reduced state of Eve conditioned on Alice having observed the outcomes $\mathbf{a}$.

The length of the secret key ${\bf k}$ obtained by processing the raw key ${\bf a}$ with an error correcting protocol and a 2-universal random function is, up to terms of order $\sqrt{N}$,  lower bounded by $H_\mathrm{min}({\bf a}|\E) - N_\mathrm{pub}$, where $H_\mathrm{min}({\bf a}|\E)$ is the min-entropy of $\bf a$ conditioned on Eve's information for the state \eqref{ccq} and $N_\mathrm{pub}$ is the length of the message published by Alice in the error-correcting phase. 
 It is shown in \cite{CK} that the length of the public message necessary for correcting Bob's errors is $N_\mathrm{pub} = N H(a|b)$, up to terms of order $\sqrt{N}$. The quantity $H(a|b)$ is the conditional Shannon entropy \cite{CK}, defined by
\begin{equation}
  H(a|b) = \sum_{a,b} -P(a,b) \log_2 P(a|b)\,,
\end{equation}
where $P(a,b)=1/N \sum_{i=1}^N \sum_{a_i,b_i} P(a_i=a,b_i=b)$ is the average probability with witch the pair of outcomes $a$ and $b$ are observed. Computing the key rate of the DIQKD protocol, thus essentially amounts to determine the min-entropy $H_\mathrm{min}({\bf a}|E)$. We show in the following how to put a bound on this quantity as a function of the estimated Bell violation $g$.

Intuitively, we want to understand how the observed Bell violation limits the predictability of Alice's outcomes $\bf a$. We start by considering the simpler case of one pair of systems ($N=1$) uncorrelated to the adversary and characterized by the joint probabilities
\begin{equation}\label{dist}
    P(ab|xy) = \tr[\rho\, A(a|x) B(b|y)]\ .
\end{equation}
If $P(a|x_\mathrm{raw})<1$ for all $a$, then the outcome of the measurement $x_\mathrm{raw}$ cannot be perfectly predicted. The degree of unpredictability of $a$ can be quantified by the probability to correctly guess $a$ \cite{meaning of min E}. This guessing probability is equal to
\begin{equation}\label{pguess}
    P_\mathrm{guess} (a) =\max_a P(a|x_\mathrm{raw})\,,
\end{equation}
since the best guess that one can make about $a$ is to output the most probable outcome.
If $P_\mathrm{guess} (a)=1$ then the outcome of the measurement $x_\mathrm{raw}$ can be predicted with certainty, while lower values for $P_\mathrm{guess} (a)$ imply less predictability.

Let $g_\mathrm{exp}=\sum_{abxy}g_{abxy}P(ab|xy)=\tr[\rho\,G]$ denote the expected quantum violation of the Bell inequality
\eqref{g} for the pair of systems described by \eqref{dist}, where
\begin{equation}
    G= \sum_{a,b,x,y} g_{abxy} A(a|x) B(b|y)\,,
\end{equation}
is the Bell operator associated to the inequality $g$ and to the measurements $A(a|x)$ and $B(b|y)$. Independently of the precise form of the state $\rho$ and of the measurement operators $A(a|x)$ and $B(b|y)$, the value of the Bell expectation $g_\mathrm{exp}$ imposes a constraint on the guessing probability \eqref{pguess}. In the case of the CHSH inequality, for instance, the following (tight) bound holds (see Appendix~A and Ref.~\cite{RND})
\begin{equation}\label{fchsh}
    P_\mathrm{guess}(a)\leq 
    \frac{1}{2} + \frac{1}{2}\sqrt{2 - \frac{g_\mathrm{exp}^2}{4} }\,,
\end{equation}
for any of the two possible values $x_\mathrm{raw}=0$ or $1$ entering in the CHSH definition \eqref{gchsh}.

More generally, let
\begin{equation}\label{f}
  P_\mathrm{guess} (a) \leq f(g_\mathrm{exp})\ ,
\end{equation}
be a bound between the guessing probability and the Bell violation, where $f$ is a concave and monotically decreasing function. Such a bound can always be obtained using the semidefinite programming (SDP) method introduced in \cite{Q set}. Indeed, the maximal value of the guessing probability $P_\mathrm{guess}(a)$ for a given value of the Bell expectation $g_\mathrm{exp}$ corresponds to the solution of the following optimization problem
\be \label{foptim}
\begin{array}{cl}
\displaystyle\max_{\rho,A,B}\quad&\tr[\rho\, A(a|x_\mathrm{raw})]\\
\text{subject to}\quad&\tr[\rho\, G]=g_\mathrm{exp}\,,
\end{array}
\ee
where the maximum is taken over all quantum states $\rho$ and measurement operators $A(a|x)$ and $B(b|y)$. 
Following reference \cite{Q set}, one can introduce a hierarchy of SDP relaxations of the problem (\ref{foptim}). The solution to any of these SDP relaxations yields an upper-bound to the optimal solution of (\ref{foptim}) and thus a bound of the form (\ref{f}), as illustrated on Figure~1 for different  Bell inequalities. The resulting function $f$ is then always concave and monotonically decreasing, as follows from the convex nature of the problem \eqref{foptim} and of its associated SDP relaxations. Note that relaxations higher in the hierarchy necessitate more computational resources but yield better upper-bounds. In the asymptotic limit, one has the guarantee that these upper-bounds will converge to the exact maximum of (\ref{foptim}), though usually a few steps in the hierarchy already give the optimal bound (this is the case for instance for the CHSH inequality).
\begin{figure}[t]
\begin{center}
\includegraphics[scale=0.3]{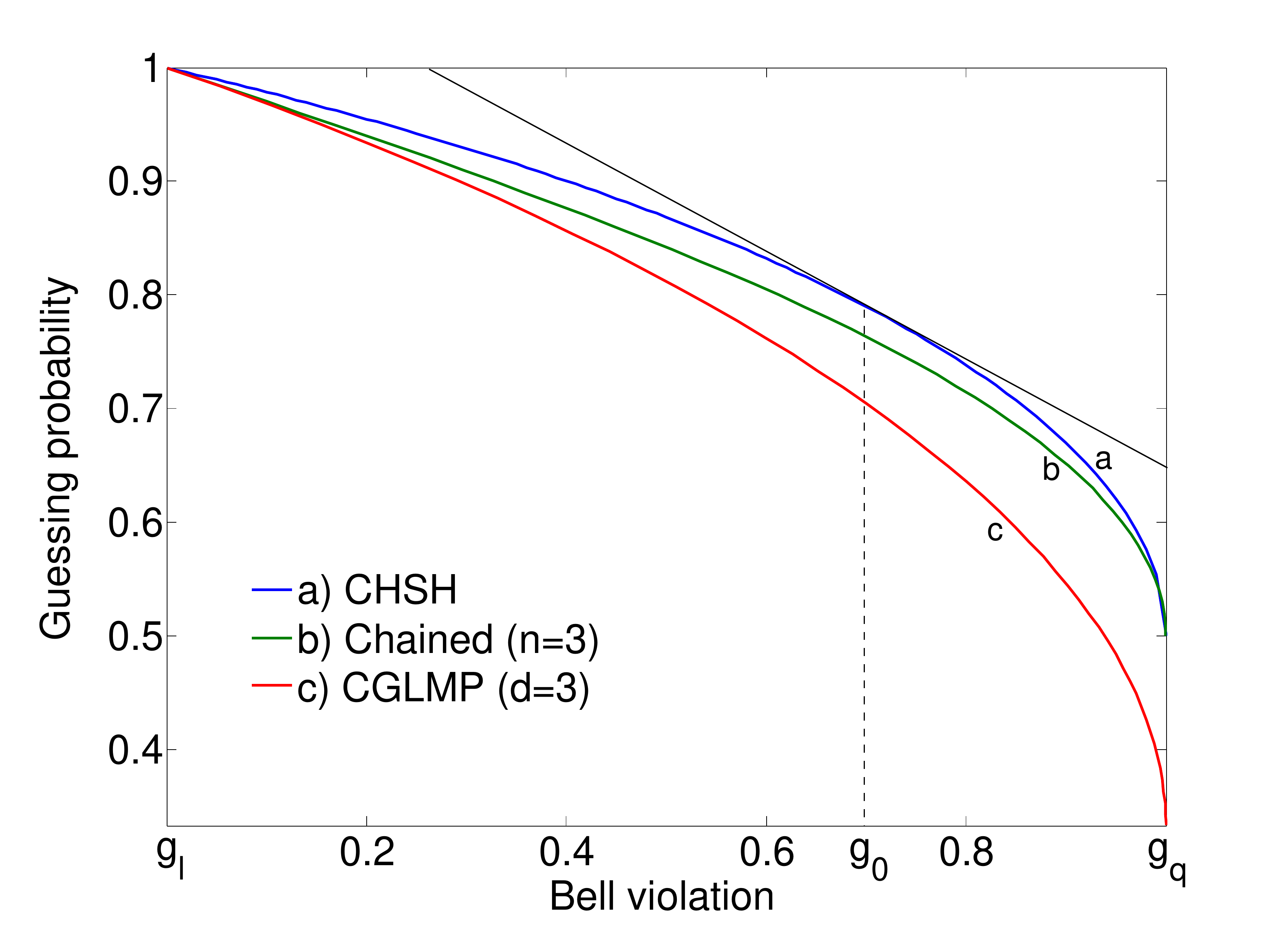}
\caption{Guessing probability $P_\mathrm{guess}(a)$ vs Bell violation $g_\mathrm{exp}$ for the CHSH inequality, the chained inequality with $n=3$ inputs~\cite{chained}, and the Collins-Gisin-Linden-Massar-Popescu (CGLMP) inequality with $d=3$ outputs~\cite{CGLMP}. Note that the symmetry of these inequalities implies that the bounds on the guessing probabilities are the same for any inputs $x_\mathrm{raw}$ entering into their definition. The horizontal scale represents the relative violation ranging from the local bound $g_{\rm loc}$ to the maximal quantum bound $g_q$. The CHSH curve is given by the function \eqref{fchsh}, the chained and CGLMP inequalities curves have been obtained by solving the problem \eqref{foptim} using the SDP relaxations introduced in \cite{Q set}. These last two curves upper-bound the optimal values by at most $O(10^{-4})$. The solid line represents a linearization of the form \eqref{linf} of the CHSH function around a point $g_0$.}
\label{fig1}
\end{center}
\end{figure}

As the function $f$ is concave, it can be upper-bounded by its linearization around any point~$g_{0}$
\begin{equation}\label{linf}
  f(g) \leq \mu(g_0) + \nu(g_0) g\ ,
\end{equation}
where $\mu(g_0)= f(g_0) -f'(g_0)g_0$, $\nu(g_0)=  f'(g_0)$.
From concavity, it also follows that
\begin{equation}\label{min}
    f(g)= \min_{g_0} \left[ \mu(g_0) + \nu(g_0) g\right] \ .
\end{equation}
The bound \eqref{f} is thus equivalent to the family of inequalities $P(a|x_\mathrm{raw}) \leq \mu(g_0) + \nu(g_0)\, g_\mathrm{exp}$
for all $a$ and $g_0$. Since these inequalities are satisfied by any quantum distribution \eqref{dist}, and thus in particular by any state $\rho$, they are equivalent to the operator inequalities
\begin{equation}\label{op ineq}
    A(a|x_\mathrm{raw})  \leq \mu(g_0)\one
    + \nu(g_0) G\ ,
\end{equation}
valid for all $a$, $g_0$, and any set of measurements $A(a|x)$ and $B(b|y)$. A proof of the bound \eqref{f} for the CHSH inequality based on such operator inequalities is given in Appendix~A. In general, the validity of any linear operator inequality of the form~\eqref{op ineq} can be established, independently of the Hilbert space dimension, using the dual formulation \cite{SIAM} of the SDP techniques introduced in \cite{Q set}.

We now move to the case of $N$ pairs of systems described by \eqref{multiP} and \eqref{ccq} and evaluate the probability with which Eve can correctly guess the raw key $\mathbf{a}$ by measuring her side information $\mathcal{E}$. Suppose thus that Eve performs some measurement $z$ on her system $\mathcal{E}$ and obtains an outcome $e$. Let $P(\mathbf{a}| \mathbf{x}_\mathrm{raw}, ez)$ denote the probability distribution of $\bf a$ conditioned on Eve's information. On average, her probability to correctly guess $\bf a$ is given by $\sum_eP(e|z)\max_{\bf a} P(\mathbf{a}|\mathbf{x}_\mathrm{raw},ez)$, and her optimal correct-guessing probability (optimized over all measurements $z$) is \cite{meaning of min E}:
\begin{equation}\label{Pguess}
    P_\mathrm{guess} ({\bf a}|\mathcal{E}) =
    \max_z \sum_e P(e|z)\max_{\bf a} P({\bf a}| {\bf x}_\mathrm{raw},ez)\ .
\end{equation}
Denote by $\rho_{\mathcal{AB}|ez}$ the $2N$-partite state prepared when Eve measures $z$ and obtains the outcome $e$ (with $\rho_{\mathcal{AB}}=\sum_e P(e|z)\rho_{\mathcal{AB}|ez}$), and write $\mathbf{A}(\mathbf{a}|\mathbf{x}_\mathrm{raw})=\prod_{i=1}^N A_i(a_i|x_\mathrm{raw})$, so that
\be
P({\bf a}| {\bf x}_\mathrm{raw},ez)=\tr\left[\rho_{\mathcal{AB}|ez} \mathbf{A}(\mathbf{a}|\mathbf{x}_\mathrm{raw})\right]\,.
\ee
Consider the following $N$-partite Bell operator
\be
\mathbf{G}(g_0)=\prod_{i=1}^N [ \mu(g_0) \one
    +\nu(g_0) G_i]\ ,
\ee
where $G_i=\sum_{a,b,x,y} g_{abxy} A_i(a_i|x_i) B_i(b_i|y_i)$.
The single-copy operator inequality \eqref{op ineq} implies that for all ${\bf a}$ and $g_0$
\begin{equation}\label{N ineq}
\mathbf{A}(\mathbf{a}|\mathbf{x}_\mathrm{raw})\leq \mathbf{G}(g_0)\,.
\end{equation}
To show this, write $A'_i= A_i(a_i|x_\mathrm{raw})$ and $G'_i= \mu(g_0) \one +\nu(g_0) G_i$. We thus want to establish that $\prod_{i=1}^N G'_i-\prod_{i=1}^N A'_i\geq 0$.
Inequality \eqref{op ineq} implies that for all $i$, $0\leq A'_i\leq G'_i$. Defining $Z_i=G'_i-A'_i\geq 0$, note then that $\prod_{i=1}^N G'_i-\prod_{i=1}^N A'_i=\prod_{i=1}^N (Z_i+A'_i)-\prod_{i=1}^N A'_i=\prod_{i=1}^N Z_i+Z_1\prod_{i=2}^N A'_i+\ldots+\prod_{i=1}^{N-1}A'_iZ_n$. Inequality \eqref{N ineq} then follows from the fact that each term in this sum is positive since it is the product of operators that are positive and, according to \eqref{commutator}, commuting.

Using inequality~\eqref{N ineq} in \eqref{Pguess}, we find
 \begin{eqnarray}
  \nonumber P_\mathrm{guess} ({\bf a}|\E)&=& \max_z\sum_e P(e|z)\, \max_{\bf a}\tr\left[\rho_{\mathcal{AB}|ez} A(\mathbf{a}|\mathbf{x}_\mathrm{raw})\right]\nonumber \\
    &\leq & \max_z \sum_e P(e|z)\, \min_{g_0} \tr\left[\rho_{\mathcal{AB}|ez}
    \mathbf{G}(g_0)\right]\,,
   \nonumber \\
    \label{E1} & \leq  & \min_{g_0}\, \tr\left[\rho_{\A\B}\,\mathbf{G}(g_0)\right]
\end{eqnarray}
where to deduce the first inequality we used, in addition to \eqref{N ineq}, the positivity of $\rho_{\mathcal{AB}|ez}$.

Note now that the quantity $\tr\left[\rho_{\A\B}\,\mathbf{G}(g_0)\right]$ is a function of the marginal distributions $P(\mathbf{ab}|\mathbf{xy})$ of Alice and Bob only and does not involve directly the system of Eve. It is shown in \cite{nonfinished}, that Alice and Bob can estimate (with high probability) this quantity from the Bell violation $g$ observed on the randomly-chosen $N_\mathrm{est}$ pairs. More precisely, Lemma 5 from reference \cite{nonfinished} implies that the inequality
\be
\tr\left[\rho_{\A\B}\,\mathbf{G}(g_0)\right]    \leq
    \left[ \mu(g_0) + \nu(g_0) g_\mathrm{est}
    + N_\mathrm{est}^{-1/4} \right]^N
\ee
holds except with probability exponentially small in $N_\mathrm{est}$.
This, \eqref{E1}, and \eqref{min} imply that
\begin{eqnarray}\label{Pguess bound}
    P_\mathrm{guess}({\bf a}|\E) \leq
  \left[ f(g^\mathrm{est}) + N_\mathrm{est}^{-1/4} \right]^N \ .
\end{eqnarray}

Finally, it is shown in \cite{meaning of min E} that the (quantum) min-entropy  $H_\mathrm{min}({\bf a}|\E)$ of a state of the form \eqref{ccq} is given by
\begin{equation}
    H_\mathrm{min}({\bf a}|\E) = -\log_2\! {P_\mathrm{guess}({\bf a}|\E)}
    \,,
\end{equation}
which implies the asymptotic secret key rate
\begin{equation}\label{rate}
  R\geq -\log_2\! {f(g_\mathrm{est})} - H(a|b) \ .
\end{equation}
This bound on the secret-key rate constitutes the main result of our work. As mentioned previously, the second term $H(a|b)$ is standard and quantifies the amount of communication needed for the error correcting phase. The non-trivial part of our bound corresponds to the first term, which quantifies the knowledge of Eve and thus the amount of privacy amplification needed to make her information arbitrarily small.

\section{Key rate of specific protocols}

We now illustrate the above formalism on two DIQKD protocols, based respectively on the chained inequality~\cite{chained,AMP} for $n=2$ and $n=3$ inputs. This inequality reads
\begin{equation}\label{gbc}
    g_\mathrm{c} = \sum_{a,b} \sum_{x=0}^{n-1} \sum_{y=x-1}^{x}(-1)^{a+b+\delta(y)} q(ab|x y) \leq 2\, ,
\end{equation}
where $a,b\in \{0,1\}$ and $x,y\in \{0,1,\ldots n-1\}$ are defined modulo $n$; the $\delta(y)$ equals one when $y=-1$ and zero otherwise. Note that for $n=2$, the chained inequality reduces to the CHSH-inequality.

In both protocols, the observed correlations $P(ab|xy)$ are obtained by measuring a two-qubit maximally entangled state $|\phi\rangle=|00\rangle+|11\rangle$ along $n$ possible directions for Alice and $n+1$ for Bob. The inputs $x_\mathrm{raw}=n-1$ and $y_\mathrm{raw}=n$ correspond to measurements in the computational basis $\{|0\rangle,|1\rangle\}$ and are used to generate the raw key. The chained inequality violation is estimated using the inputs $x,y\in\{0,\ldots,n-1\}$ and the corresponding measurement directions are set-up to obtain the maximal violation of the chained inequality given by $2\sqrt{2}$ and $3\sqrt{3}$ for the cases $n=2$ and $n=3$, respectively. For the sake of illustration, let us assume that the effect of the noise in the protocol amounts to the distribution of an entangled state $v |\phi\rangle\langle\phi|+(1-v)\one/4$ of visibility $v$. The conditional Shannon entropy 
 $H(a|b)$ is then equal to $h[(1-v)/2]$, where $h(x)=-x\log_2(x)-(1-x)\log_2(1-x)$ is the binary entropy, and the observed Bell violations are equal to
 $g=2\sqrt{2}v$ and $g=3\sqrt{3}v$. For the CHSH inequality, we then obtain using \eqref{fchsh} and \eqref{rate} the key-rate
\be
R\geq 1-\log_2 \left[1+\sqrt{2-2v^2}\right]-h\left[(1-v)/2\right]\,.
\ee
The value of the visibility such that this bound is equal to zero corresponds to a quantum-bit-error rate (QBER) of $5\%$. The key-rate for the chained inequality for $n=3$ is plotted in Figure~ 2, based on the SDP bound of Figure~1. The critical visibility corresponds to a QBER of $7.5\%$, comparable to those obtained for standard QKD. Numerical evidence suggests that the chained inequalities for a larger number of settings, $n>3$, provide worse lower bounds on the key rate.

The fact that our approach can be applied to protocols based on arbitrary Bell inequalities is particularly interesting from a practical point of view. As shown in Figure~2, using inequalities other than CHSH may lead to better key rates in the presence of noise. It could also be very useful to improve the resistance of DIQDK protocols to photon detection inefficiencies \cite{ampli}, since relevant improvements over CHSH can be obtained in realistic situations \cite{vertesi}.

\begin{figure}[t]
\begin{center}
\includegraphics[scale=0.3]{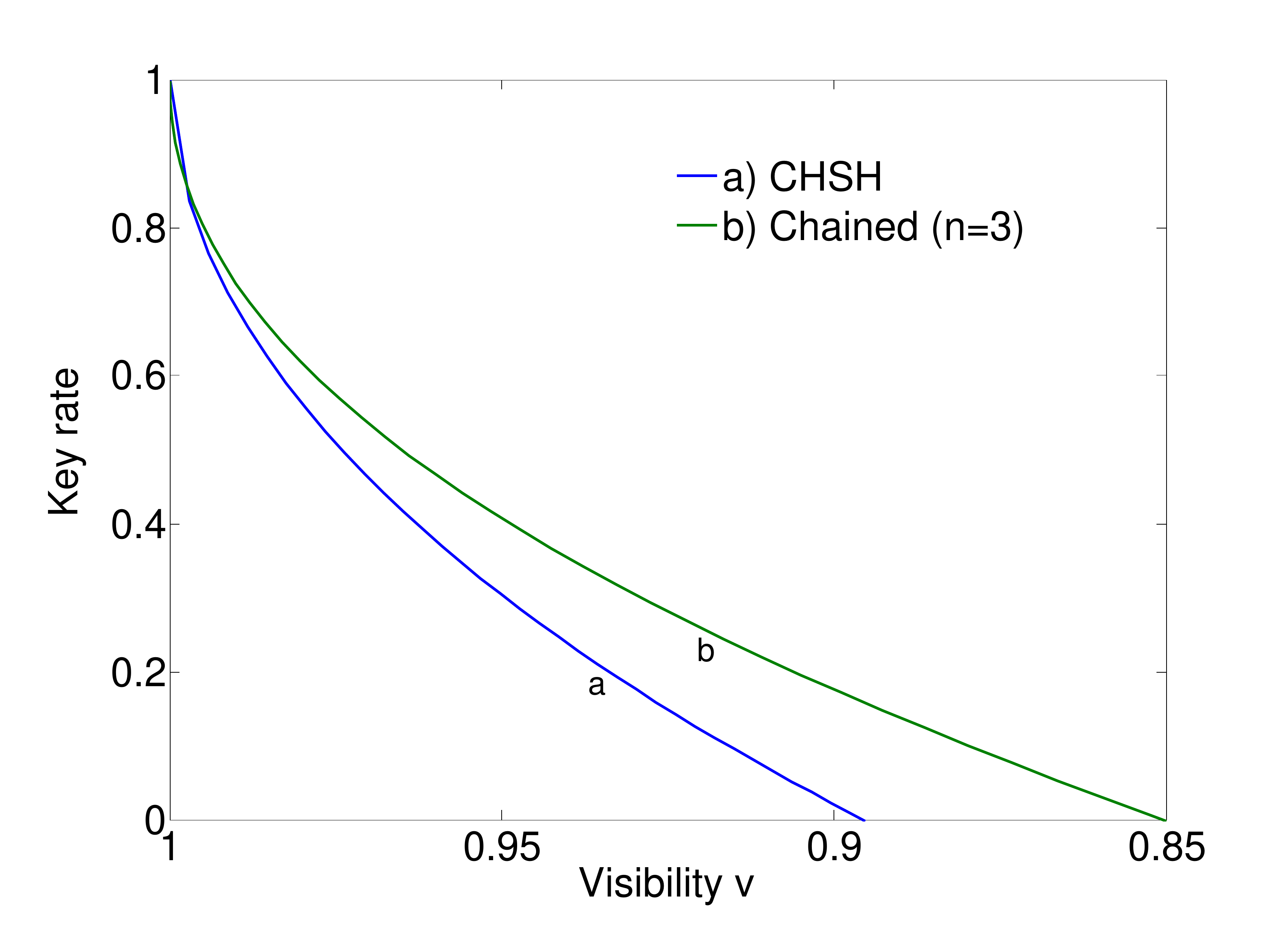}
\caption{Key rate vs visibility for the CHSH inequality and the chained inequality with 3 inputs. The key rate is given by the  formula \eqref{rate} where the function $f$ is given by \eqref{fchsh} for the CHSH inequality and has been obtained by solving the problem \eqref{foptim} using the SDP relaxations introduced in \cite{Q set} for the chained inequality (see Figure~1). Interestingly for the particular type of noise illustrated here, the chained inequality leads to better key rate than CHSH.}
\label{fig2}
\end{center}
\end{figure}

\section{Conclusion}
We have shown how to compute a bound on the key rate of a large
class of DIQKD protocols\footnote{It is easy to see that
our security proof can also be adapted to cover the less efficient
protocols introduced in \cite{AGM} and~\cite{co}, or
protocols with pre-processing of the raw key~\cite{AMP}.}. Our
approach is based on a fundamental relation between the amount by
which two quantum systems violate a Bell inequality and the
unpredictability of their local measurement outcomes, as
illustrated in Figure~1. A similar relation has been used in the
context of device-independent randomness generation~\cite{RND}.

To derive our security proof, we have used the fact that the
behavior of $N$ uses of the quantum devices is represented by
probabilities of the form \eqref{multiP} with measurement
operators satisfying the commutation relations \eqref{commutator}.
These commutation relations can be satisfied in a physical
realization in which $N$ pairs of separated and non-interacting
devices are used to generate the $N$ symbols of the raw key. If
necessary, these commutation relations can be enforced by shielding the devices in such a
way that no communication between them occurs during the
measurement process. Note that if the provider of the quantum
apparatuses is untrusted, shielding of the devices is anyway
required to guarantee that they do not send unwanted information
to the adversary. Admittedly, a realization requiring $N$
different devices for the generation of $N$ raw-key symbols is
impractical. Our results nevertheless show that secure fully
device-independent QDK with key rates comparable to those of
traditional QKD is in principle possible.

In a more realistic implementation, the raw key is generated by repeatedly performing measurements  on a single pair of devices. In such a sequential implementation, the description provided by Eqs. \eqref{multiP} and \eqref{commutator} corresponds to the assumption that the functioning of the measuring devices does not depend on an internal memory storing the quantum states and measurement results obtained at previous steps. While it would be desirable to extend our security proof to cover such possible memory effects, it may be reasonable to expect our no-memory condition to be satisfied in a variety of practical setups. After all, this no-memory condition is assumed in standard QKD, where the description of the devices fits in the formalism of Eqs.~\eqref{multiP}, \eqref{commutator2}, \eqref{commutator}.  But while we make here no assumptions at all on the measurement operators $A_i(a_i|x_i)$, $B_i(b_i|y_i)$ (nor on the Hilbert spaces on which they are defined), in standard QKD one assumes that these measuring operators have a fixed and known value, which is identical for all $i$ --- an idealized assumption difficult to verify in practice. From this perspective, the DIQKD model considered here clearly represents a \emph{relaxation} of standard QKD, and thus can only be more secure.

Note that the no-memory assumption allows for devices whose behaviour may vary with time (as implied by the dependence of $A_{i}(a_i|x_i)$ on the subindex $i$), it only excludes, e.g., that the response of the devices at step $j$ depends on the particular measurement input at step $j-k$. Such kind of memory effects could arguably be excluded, for instance if no explicit memory has been introduced in the devices or if an ``initialization" procedure is performed before every measurement based on an estimation of the apparatus memory characteristics. It may thus be legitimate to assume for particular implementations that no imperfections, failures, or implementation weaknesses would introduce detrimental memory effects (even though imperfections could be exploited in other ways by an eavesdropper). From this perspective, our work contributes to narrow the gap between theoretical security proofs and practical realizations of QKD.

{\it Note added after completion of this work:} results closely related to the ones presented here have been obtained independently in  Ref.~\cite{hr}.

\section*{Appendix: Local randomness vs CHSH violation}

Let $P(ab|xy)$ with $a,b,x,y\in\{0,1\}$ be a quantum distribution of the form~\eqref{dist} and let $g=\sum_{a,b,x,y}\linebreak[1](-1)^{a+b+xy} P(ab|xy)$ be the corresponding CHSH expectation.
We establish here (see also \cite{RND}) that
\be \label{app1}
P(a|x)\leq \frac{1}{2}+\frac{1}{2}\sqrt{2-\frac{g^2}{4}}
\ee
for all $a,x\in\{0,1\}$, which implies inequality \eqref{fchsh}. We consider only the case $a=0$ and $x=0$ (the argument applies by symmetry to the other cases as well).

Let $G=\sum_{a,b,x,y}(-1)^ {a+b+xy}A(a|X)B(b|y)$. Following the discussion after Eq.~\eqref{linf}, inequality~\eqref{app1} is equivalent for $g_0\in[2,2\sqrt{2}[$ to the series of operator inequalities
\be \label{app3b}
A(0|0)\leq \frac{1}{2}+\frac{1}{\sqrt{2-\frac{g_0^2}{4}}}-\frac{g_0}{8\sqrt{2-\frac{g_0^2}{4}}} G\,,
\ee
since $f'(g)=-g/(8\sqrt{2-g^2/4})$. By increasing the dimension of the Hilbert space, we can always take the measurement operators $A(a|x)$ and $B(b|y)$ to be projection operators. Define then operators $A_x=A(a=0|x)-A(a=1|x)$ and $B_y=B(b=0|y)-B(b=1|y)$. It is easily verified that these new operators are hermitian and satisfy $A_x^2=\one$ and $B_y^2=\one$. In term of these operators we can rewrite inequality \eqref{app3b} as
\be \label{app3}
\frac{1}{2}+\frac{1}{2}A_0 \leq \frac{1}{2}+\frac{1}{\sqrt{2-\frac{g_0^2}{4}}}-\frac{g_0}{8\sqrt{2-\frac{g_0^2}{4}}} G\,,
\ee
where $G=A_0B_0+A_0B_1+A_1B_0-A_1B_1$
We now prove this operator inequality. For this, let $\alpha=1/(\sqrt{8-g_0^2})$,  $\gamma_1=\sqrt{\alpha}/4$, $\gamma_2=-g_0\sqrt{\alpha}/8$, $\gamma_3=g_0/(16\sqrt{\alpha})$, $\gamma_4=1/(8\sqrt{\alpha})$, and $\gamma_5=(1-g_0^2/4)\sqrt{\alpha}/4$, and define the following four operators
\begin{eqnarray}
O_1&=&-2\gamma_1 A_2 -\gamma_2 (B_1-B_2)+\gamma_3 (A_2B_1+A_2B_2)\nonumber\\
O_2&=&-2\gamma_2-2\gamma_3 A_1 +\gamma_4 (B_1 +B_2)\nonumber\\
&&\qquad-\gamma_1 (A_1B_1+A_1B_2)+\gamma_5 (A_2B_1-A_2B_2)\nonumber\\
O_3&=&-\gamma_4 (B_1 -B_2)+\gamma_1 (A_1B_1-A_1B_2)\nonumber\\
&&\qquad-\gamma_5 (A_2B_1+A_2 B_2)\nonumber\\
O_4&=&2\gamma_4-2\gamma_5 A_1 +\gamma_2 (B_1 + B_2)-\gamma_3 (A_2B_1- A_2B_2)\nonumber\,,
\end{eqnarray}
Using the fact that  $A_x^2=\one$, $B_y^2=\one$, and $[A_x,B_y]=0$, the following algebraic idendity is easily verified
\be
\sum_i O_i^\dagger O_i=-\frac{1}{2}A_0\otimes\one+\frac{1}{\sqrt{2-\frac{g_0^2}{4}}}-\frac{g_0}{8\sqrt{2-\frac{g_0^2}{4}}} G\,.
\ee
Note now that since the left hand side is a sum of square, it is necessarily positive semidefinite, i.e., $\sum_i O_i^\dagger O_i\geq 0$, which immediately implies \eqref{app3}. Note that we have established inequality \eqref{app1} only for $g_0\in[2,2\sqrt{2}[$. The bound for $g_0=2\sqrt{2}$ follows from the fact that the function $f(g)$ corresponding to the right-hand side of \eqref{app1} is concave and monotonically decreasing and hence $f(2\sqrt{2})\leq \lim_{\epsilon\to 0}f(2\sqrt{2}-\epsilon)=1/2$.

Finally, we show that inequality \eqref{app1} is optimal, i.e.,  that there exists quantum states and operators that saturate the inequality. Consider the two-qubit state $\cos\theta |00\rangle+\sin\theta|11\rangle$, and the measurement operators $A_0=\sigma_z\otimes\one$, $A_1=\sigma_x\otimes\one$, $B_0=\one\otimes\cos \phi \sigma_z+\sin\phi\sigma_x$, and $B_1=\one\otimes\cos \phi \sigma_z-\sin\phi\sigma_x$, where $\tan\phi=\sin 2\theta$ and $2\sqrt{1+\sin^2(2\theta)}=g$. It is straightforward to see that the corresponding quantum probabilities $P(ab|xy)$ saturate the inequality \eqref{app1} for all values of $g\in[2,2\sqrt{2}]$.

\paragraph{Acknowledgments.}
This work is supported by the Spanish MEC/MINCIN projects QTIT (FIS2007-60182) and QOIT (Consolider Ingenio 2010), EU Integrated Project Q-Essence and ERC Starting Grant PERCENT, Caixa Manresa, Generalitat de Catalunya, and the Brussels-Capital region through a BB2B grant.


\end{document}